\documentclass{article}

\usepackage{amsmath,amssymb,amsfonts}
\usepackage{a4wide}
\usepackage{color}
\usepackage{bm}
\usepackage{graphicx}
\usepackage{euscript}
\usepackage{ulem}

\definecolor{gray}{gray}{0.6}

\date{\today}
\title{Nonequilibrium physics in integrable systems and spin-flip non-invariant conserved quantities}
\author{Chihiro Matsui \\[3ex]
{\it Graduate School of Mathematical Sciences, The University of Tokyo} \\
{\it 3-8-1, Komaba, Meguro-ku, 153-8914 Tokyo, Japan}}

\begin{document}
\maketitle

\begin{center}
{\bf Abstract}
\end{center}
\bigskip
{\small
Recently found spin-flip non-invariant (SFNI) conserved quantities play important roles in discussing nonequilibrium physics of the $XXZ$ model. The representative examples are the generalized Gibbs ensemble (GGE) and the ballistic transport of the spin current. In spite of big progress in understanding nonequilibrium physics of integrable systems, the general framework to determine a minimal complete set of conserved quantities which describes the long-time steady state has not yet been found. This paper shows that the GGE of the gapless $XXZ$ model consists of functionally independent conserved quantities rather than linearly independent. At the same time, the physical meaning of SFNI conserved quantities is provided. We also discuss that there exist ballistic channels of the spin current supported by non-quasilocal conserved quantities. The saturation of the lower bound for the Drude weight by quasilocal
conserved quantities reads the linear dependence of non-quasilocal conserved quantities on quasilocal ones. We show that their (generalized) linearly dependence relation is consistent with the statement that the GGE consists of functionally independent conserved quantities without containing all linearly independent conserved quantities. 
}

\section{Introduction}
Recently, it has been found for the $XXZ$ model that the spin-flip non-invariant (SFNI) conserved quantities exist~\cite{bib:PI13}, although the model itself is spin-flip symmetric. Here we use the word ``spin-flip''as the operation to exchange the roles of an up spin and a down spin. The conserved quantities of integrable systems are obtained from the parameter expansion of the transfer matrix that consists of the ordered product of the Lax operators. The Lax operator is defined on the tensor product of the auxiliary space and the local quantum space. The quantum space is the physical space to be chosen as the spin-$\frac{1}{2}$ representation for the $XXZ$ model, whereas the auxiliary space does not appear in physical quantities and can be arbitrarily chosen. The spin-flip non-invariance of conserved quantities occurs when we choose the complex-spin representations for the auxiliary space of the Lax operators. 
Some of SFNI conserved quantities possess quasilocality~\cite{bib:P14} and, consequently, extensivity~\cite{bib:IMPZ16}, which is the expected property as thermodynamic variables. Indeed, the SFNI conserved quantities play quite important roles in discussing nonequilibrium physics, as we will see below. 

The existence of macroscopic number of conserved quantities brings interesting nonequilibrium phenomena. The representative example is non-thermalization of integrable systems. Although ``the eigenstate thermalization hypothesis'' (ETH)~\cite{bib:RDO08} succeeded to explain the mechanism why isolated systems thermalize, ETH is no more true for integrable systems. Instead, ``the generalized Gibbs ensemble'' (GGE)~\cite{bib:RDYO07} has been proposed to describe the steady state which an integrable system approaches in the long-time limit. The GGE is the generalization of the Gibbs ensemble 
that consists of a macroscopic number of conserved quantities $Q_r$: 
\begin{equation} \label{eq:GGE}
 \rho_{\rm GGE} = Z^{-1} e^{-\sum_r \beta_r Q_r}, \quad
  Z = {\rm tr}\,e^{-\sum_r \beta_r Q_r}. 
\end{equation}
This means that the existence of as many conserved quantities as the order of the system-size strongly restricts relaxation processes of the system. The question which has been discussed is, among infinitely many conserved quantities existing for integrable systems, which conserved quantities form a minimal complete set to constitute the GGE. 
The description of the steady state of the $XXZ$ model by the GGE is well-studied through ``the string-charge duality''~\cite{bib:IQNB16}. As a Bethe-ansatz solvable model, the steady state is characterized by the Bethe string densities in the thermodynamic limit~\cite{bib:C16}. The string-charge duality provides the correspondence between the expectation values of conserved quantities on the initial state and the Bethe string densities for the steady state. Therefore, a set of conserved quantitites which completely determines the Bethe string densities of the steady state is considered to constitute the GGE. 
There are several trials to this direction~\cite{bib:INWCEP15, bib:IQNB16, bib:LCN17, bib:WNBFRC14}. The complete GGE has been heuristically constructed by using the conserved quantities associated with (half-)integer spins and one complex spin~\cite{bib:LCN17}. In this paper, we explain why adding one complex spin conserved quantity completes the GGE from the viewpoint of independence of conserved quantities. 

Another interesting nonequilibrium phenomena brought by many conserved quantities of integrable systems is non-vanishing currents. Under the presence of many conserved quantities, ballistic transport of currents and hence the finite Drude weight has been predicted~\cite{bib:SPA09, bib:ZNP97}. In the context of the linear response theory, the Drude weight is evaluated by the current-current correlation whose lower bound is given by the overlap with an ``orthogonal set'' of conserved quantities~\cite{bib:M69, bib:PI13, bib:S71, bib:ZNP97}: 
\begin{equation}
 D(\beta) \geq \lim_{N \to \infty} \frac{\beta}{2N} \sum_k \frac{|\langle J,Q_k \rangle_{\beta}|^2}{|| Q_k ||^2_{\beta}},\quad
  \langle Q_j,Q_k \rangle_{\beta} = \delta_{j,k} || Q_k ||_{\beta}^2. 
\end{equation}
Since each conserved quantity, if it has finite overlap with the current, supports a ballistic channel, ``a complete set'' of conserved quantities covers all ballistic channels and saturates the above lower bound. 
As a spin-flip anti-symmetric (SFAS) operator, the spin current operator has overlap only with the spin-flip non-symmetric conserved quantities~\cite{bib:PI13}. The saturation of the lower bound for the Drude weight by quasilocal SFNI conserved quantities has been suggested in comparison with the result obtained from the thermodynamic Bethe ansatz~\cite{bib:Z99}, although, in our analysis, non-quasilocal conserved quantities also provide ballistic channels of the spin current in the thermodynamic limit at high temperature. Actually, the improvement of the lower bound occurs only when the conserved quantities are linearly independent, which is checked by decomposing an operator into conserved quantities~\cite{bib:S71}. This implies that non-quasilocal SFNI conserved quantities are the (generalized) linear combination of quasilocal ones in the thermodynamic limit. 
We show that a subsequently obtained relation from the generalized linear combination is consistent with the statement that conserved quantities constituting the GGE of the gapless $XXZ$ are functionally independent rather than linearly independent. 

The paper is organized as follows. We first review the construction of conserved quantities of the $XXZ$ model including SFNI ones. The complex spin representations of $\mathfrak{sl}_q(2)$ are also explained. The locality and extensivity of SFNI conserved quantities is discussed from the large-volume analysis of the operator inner product and norm. After then, two important applications are discussed. The first one is the GGE which we show consists of a set of functionally independent conserved quantities. The physical meaning of SFNI conserved quantities is also provided. In the next section, we discuss the non-vanishing spin current. We show that the non-quasilocal conserved quantities provide the ballistic channels. The saturation of the lower bound by the quasilocal conserved quantities indicates that the non-quasilocal conserved quantities are linearly dependent on the quasilocal ones in the thermodynamic limit, which we show is
consistent with the functional dependence of conserved quantities in the GGE. 

\section{Conserved quantities without spin-flip invariance}
The $XXZ$ model is known to be integrable, which has as many conserved quantities as the order of the system size $N$. Many conserved quantities arise from commuting transfer matrices. 
In this section, we review how conserved quantities of the $XXZ$ model are constructed. Then we discuss how the SFNI conserved quantities are obtained from the spin-flip symmetric $XXZ$ model. Quasilocality and extensivity of the conserved quantities is also discussed. 

\subsection{The model}
Let us consider the $XXZ$ model defined on the Hilbert space given by the tensor product $\mathcal{H} = \prod_{n=1}^N \otimes h_n$. The Hamiltonian is given by 
\begin{equation} \label{eq:Hamiltonian}
 H = \sum_{n=1}^N (\mathsf{S}_n^x \mathsf{S}_{n+1}^x + \mathsf{S}_n^y \mathsf{S}_{n+1}^y + \cos \gamma \mathsf{S}_n^z \mathsf{S}_{n+1}^z)
\end{equation}
where $\gamma$ determines the anisotropy of the model. The model shows different physics depending on $\gamma$ by showing the gapped energy spectrum for pure imaginary $\gamma$, whereas the gapless energy spectrum for real $\gamma$. The spin operators $\mathsf{S}_n^{\alpha}$ ($\alpha \in \{x,y,z\}$) are ultralocal operators in the sense that they nontrivially act only on the $n$th quantum space $h_n$: 
\begin{equation}
 \mathsf{S}_n^{\alpha} = \bm{1} \otimes \cdots \otimes \bm{1} \otimes \underset{n}{\mathsf{S}^{\alpha}} \otimes \bm{1} \otimes \cdots \otimes \bm{1}. 
\end{equation}
We impose the periodic boundary condition so that $n \equiv n + N$. 
The defining ultralocal algebra for the spin operators $\mathsf{S}^{\alpha}$ is the $\mathfrak{sl}(2)$ commutation relations:
\begin{equation}
 [\mathsf{S}_m^{\alpha},\,\mathsf{S}_n^{\beta}] = i\epsilon_{\alpha\beta\gamma} \mathsf{S}_n^{\gamma} \delta_{mn}
\end{equation}
where $\epsilon_{\alpha\beta\gamma}$ is a completely antisymmetric tensor $\epsilon_{123} = 1$. The non-trivial finite-dimensional representations for $\mathsf{S}^{\alpha}$ are realized in $\mathbb{C}^{2s+1}$ labeled by positive (half-)integers $s = \frac{1}{2},1,\dots$. For the $XXZ$ model, we choose the smallest nontrivial representations realized by $s = \frac{1}{2}$: 
\begin{equation}
 \mathsf{S}^1 = \begin{pmatrix} 0 & \frac{1}{2} \\ \frac{1}{2} & 0 \end{pmatrix}, \quad
 \mathsf{S}^2 = \begin{pmatrix} 0 & -\frac{i}{2} \\ \frac{i}{2} & 0 \end{pmatrix}, \quad
 \mathsf{S}^3 = \begin{pmatrix} \frac{1}{2} & 0 \\ 0 & -\frac{1}{2} \end{pmatrix}. 
\end{equation}
Although written in terms of the $\mathfrak{sl}(2)$ spin operators, the model is more related to the $q( = e^{i\gamma})$-deformed $\mathfrak{sl}(2)$ algebra, denoted by $\mathfrak{sl}_q(2)$, due to the anisotropy. The $q$-deformed spin operators $S^{\pm}, q^{S^z}$ define the $\mathfrak{sl}_q(2)$ algebra by the relations 
\begin{equation}
 q^{S^z} S^{\pm} = q^{\pm 1} S^{\pm} q^{S^z}, \quad
  [S^+,\,S^-] = \frac{(q^{S^z})^2 - (q^{S^z})^{-2}}{q - q^{-1}}, 
\end{equation}
which are reduced to the normal $\mathfrak{sl}(2)$ relations at the $q \to 1$ limit by identifying $S^z = \mathsf{S}^3, S^{\pm} = \mathsf{S}^1 \pm i\mathsf{S}^2$. 
The finite-dimensional representations are realized in $\mathbb{C}^{2s+1}$ as in the $\mathfrak{sl}(2)$ case for generic $q$. The special care is required for $q$ at the root of unity, as is explained later. 

The Lax operator is defined in the tensor product of the auxiliary space $V$ and the local quantum space $h_n$. Let $V$ be $\mathbb{C}^2$. Then the Lax operator acting in $V \otimes h_n$ is written as the $2 \times 2$-matrix in the auxiliary space: 
\begin{equation} \label{eq:Lax_half}
 L_{a,n}(\lambda) = 
  \begin{pmatrix}
   \sinh(\lambda + i\gamma S_n^z) & i\sin\gamma \cdot S_n^- \\ i\sin\gamma \cdot S_n^+ & \sinh(\lambda - i\gamma S_n^z)
  \end{pmatrix}_a
\end{equation}
with the entries being operators in the quantum space $h_n$. 
The Lax operator satisfies the $RLL$ relation in $V_1 \otimes V_2 \otimes h_n$: 
\begin{equation}
 R_{a_1,a_2}(\lambda - \mu) L_{a_1,n}(\lambda) L_{a_2,n}(\mu)
  = L_{a_2,n}(\mu) L_{a_1,n}(\lambda) R_{a_1,a_2}(\lambda - \mu), 
\end{equation}
where $R_{a_1,a_2}$ is the $R$-matrix which nontrivially acts on $V_1 \otimes V_2$ as 
\begin{equation}
 R_{a_1,a_2}(\lambda) = L_{a_1,a_2}\left( \lambda + \frac{i\gamma}{2} \right). 
\end{equation}
We call the ordered product of the Lax operators the monodromy matrix: 
\begin{equation}
 \mathsf{T}_{a}(\lambda) = L_{a,N}(\lambda) \cdots L_{a,1}(\lambda). 
\end{equation}
The monodromy matrix satisfies the $R\mathsf{T}\mathsf{T}$ relation in the tensor product $V \otimes \prod_{n=1}^N \otimes h_n$: 
\begin{equation}
 R_{a_1,a_2}(\lambda - \mu) \mathsf{T}_{a_1}(\lambda) \mathsf{T}_{a_2}(\mu)
  = \mathsf{T}_{a_2}(\mu) \mathsf{T}_{a_1}(\lambda) R_{a_1,a_2}(\lambda - \mu) 
\end{equation}
as a result of the $RLL$ relation. 
The transfer matrix is defined by the trace of the monodromy matrix over the auxiliary space: 
\begin{equation}
 T(\lambda) = {\rm tr}_a\mathsf{T}_a(\lambda) 
\end{equation}
that is commuting for different parameters $\lambda$ due to the $R\mathsf{T}\mathsf{T}$ relation: 
\begin{equation}
 [T(\lambda),\,T(\mu)] = 0. 
\end{equation}

The transfer matrix allows the $\theta$-expansion in the $N \to \infty$ limit: 
\begin{equation}
 T(i\theta) = i^N \sum_{r=0}^{\infty} \frac{(\theta - \theta_0)^r}{r!} Q_r(\theta_0). 
\end{equation}
Since the Hamiltonian \eqref{eq:Hamiltonian}, which is explicitly written as the logarithmic derivative of the transfer matrix: 
\begin{equation}
 H \propto \frac{d}{d\lambda} \log T(\lambda) \Big|_{\lambda = \frac{i\gamma}{2}}
\end{equation}
belongs to this family, we call commuting operators $Q_r(\lambda_0)$ conserved quantities. Note that the total spin operator $S^z$ does not belong to this family since it is a SFAS operator. 

The auxiliary space is straightforwardly generalized to $V = \mathbb{C}^{2s+1}$ for (half-)integers $s$. Let $|r \rangle$ ($r = 0,\dots, 2s$) be a natural basis in $\mathbb{C}^{2s+1}$: 
\begin{equation}
 |r \rangle = (0 \quad \cdots \quad 0 \quad \underset{r}{1} \quad 0 \quad \cdots \quad 0)^{\rm T}. 
\end{equation}
The Lax operators for $V = \mathbb{C}^{2s+1}$ is given by the same expression for the $V = \mathbb{C}^2$ case \eqref{eq:Lax_half} but by inserting the (half-)integer spin-$s$ representations into each spin operator $S_a^{\pm}, q^{S_a^z}$: 
\begin{equation} \label{eq:spin-s_rep}
 S_a^+ = \sum_{r=0}^{2s-1} \frac{\sin(\gamma (r+1))}{\sin\gamma} |r \rangle \langle r+1|, \quad
  S_a^- = \sum_{r=0}^{2s-1} \frac{\sin(\gamma (2s-r))}{\sin\gamma} |r+1 \rangle \langle r|,\quad
  q^{S_a^z} = \sum_{r=0}^{2s} e^{i\gamma (s-r)} |r \rangle \langle r|. 
\end{equation}
Note that here we consider only generic $q$. (The case for $q$ at the root of unity is discussed in the next subsection. ) 
The (half-)integer spin-$s$ representations are the highest weight representations $S^+ |0 \rangle = 0$ and finite dimensional representations $S^- |2s \rangle = 0$. Under the choice of (half-)integer $s$, the action of the transpose exchanges $S^{\pm}$-operators :  
\begin{equation} \label{eq:spin_flip}
 (S^{\pm})^{\rm T} = S^{\mp}. 
\end{equation}
Let $V_2$ be $V_2 = \mathbb{C}^{2s+1}$ by keeping $V_1$ as $V_1 = \mathbb{C}^2$. Remarkably, the product of the Lax operators still satisfy the $RLL$ relation in $V_1 \otimes V_2 \otimes h_n$~\cite{bib:KRS81}: 
\begin{equation}
 R^{( \frac{1}{2},s)}_{a_1,a_2}\left( \lambda - \mu \right) L^{(\frac{1}{2})}_{a_1,n}\left( \lambda \right) L^{(s)}_{a_2,n}(\mu)
  = L^{(s)}_{a_2,n}(\mu) L^{(\frac{1}{2})}_{a_1,n}\left( \lambda \right) R^{(\frac{1}{2},s)}_{a_1,a_2}\left( \lambda - \mu \right) 
\end{equation}
leading to the commuting transfer matrices: 
\begin{equation}
 [T^{(\frac{1}{2})}(\lambda),\,T^{(s)}(\mu)] = 0, \quad
  T^{(s)}(\mu) = {\rm tr}_a \left( L^{(s)}_{a,N}(\mu) \cdots L^{(s)}_{a,1}(\mu) \right). 
\end{equation}
The above commutativity of the transfer matrices provides another family of conserved quantities. Especially at large $N$, the transfer matrices admit the $\theta$-expansion: 
\begin{equation}
 T^{(s)}(i\theta) = i^N \sum_{r=0}^{\infty} \frac{(\theta - \theta_0)^r}{r!} Q^{(s)}_r(\theta_0), 
\end{equation}
where the conserved quantities are obtained as the coefficients $Q^{(s)}_r(\theta_0)$.

\subsection{Complex spin representations and associated conserved quantitites}
The $\mathfrak{sl}_q(2)$ spin operators admit arbitrary spin-$s \in \mathbb{C}$ representations besides the (half-)integer spin representations. Let $q$ be generic. The complex spin-$s$ representations are still the highest weight representations $S^+|0 \rangle = 0$ but infinite dimensional: 
\begin{equation} \label{eq:infinite_rep}
  S_a^+ = \sum_{r=0}^{\infty} \frac{\sin(\gamma (r+1))}{\sin\gamma} |r \rangle \langle r+1|, \quad
  S_a^- = \sum_{r=0}^{\infty} \frac{\sin(\gamma (2s-r))}{\sin\gamma} |r+1 \rangle \langle r|,\quad
  q^{S_a^z} = \sum_{r=0}^{\infty} e^{i\gamma (s-r)} |r \rangle \langle r|. 
\end{equation}
The truncation to the finite-dimensional irreducible representations occurs for the (half-)integers $s$ case which we discussed in the previous subsection. 
When $q$ is at the root of unity $q = e^{i\pi \frac{m}{l}}$ ($\gamma = \pi \frac{m}{l}$) where $m$ and $l$ are coprime, we are able to take $l \times l$-dimensional irreducible representations 
\begin{equation} \label{eq:finite_rep}
 S_a^+ = \sum_{r=0}^{l-2} \frac{\sin(\gamma (r+1))}{\sin\gamma} |r \rangle \langle r+1|, \quad
  S_a^- = \sum_{r=0}^{l-2} \frac{\sin(\gamma (2s-r))}{\sin\gamma} |r+1 \rangle \langle r|,\quad
  q^{S_a^z} = \sum_{r=0}^{l-1} e^{i\gamma (s-r)} |r \rangle \langle r|, 
\end{equation}
due to the existence of extra centers $(S^{\pm})^l$ and $(q^{S^z})^l$~\cite{bib:GRS05}. The action of the transpose does not exchange the spin operators $(S^{\pm})^{\rm T} \neq S^{\mp}$ under these representations but the highest weight vector and the lowest weight vector exchanges their roles. There also exist the other $l \times l$-dimensional irreducible representations which are cyclic or semi-cyclic. For (half-)integers $s \leq \frac{l-1}{2}$, there exist the $(2s+1) \times (2s+1)$-dimensional representations which are similar to the generic $q$ case given in \eqref{eq:spin-s_rep}. 
Note that the finite-dimensional representations associated with arbitrary complex spin-$s$ never exist in the gapped regime nor the isotropic point. 

For the Lax operator with the complex-spin auxiliary space, it is reasonable to consider $s$ as a parameter~\cite{bib:P14}:
\begin{equation}
 L_{a,n}(\lambda,s) = 
  \begin{pmatrix}
   \sinh(\lambda + i\gamma S_a^z(s)) & \sin\gamma \cdot S_a^-(s) \\ \sin\gamma \cdot S_a^+(s) & \sinh(\lambda - i\gamma S_a^z(s))
  \end{pmatrix}_n 
\end{equation}
which coincides with the Lax operator of the (half-)integer $s$ just by replacing the spin operators with those of the complex-spin representations. 
Then the transfer matrix: 
\begin{equation} \label{eq:TM_complex_s}
 T(\lambda,s) = {\rm tr}_a (L_{a,N}(\lambda,s) \cdots L_{a,1}(\lambda,s))
\end{equation}
admits the $s$-expansion as well as the $\theta (=\frac{\lambda}{i})$-expansion at large $N$: 
\begin{equation} \label{eq:TM_expansion}
T(i\theta,s) = i^N \sum_{r,r'=0}^{\infty} \frac{(\theta - \theta_0)^r}{r!} \frac{(s - s_0)^{r'}}{r'!} Q_{r,r'}(\theta_0,s_0)
\end{equation}
producing a two-parameter family of conserved quantities $Q_{r,r'}(\theta_0,s_0)$~\cite{bib:PPSA14, bib:P14}. 

Since the Hamiltonian \eqref{eq:Hamiltonian} is given by the logarithmic derivative of the transfer matrix, we are motivated to introduce another series of conserved quantities $H_{r,r'}$ obtained by the logarithmic derivatives of the transfer matrix: 
\begin{equation}
	H_{r,r'}(\lambda_0,s_0) = \partial_{\lambda}^r \partial_s^{r'} \log T(\lambda,s) \Big|_{\lambda = \lambda_0, s=s_0}. 
\end{equation}
These are more natural definition of the conserved quantities since they are extensive, i.e. their expectation values are proportional to the system size for large $N$, as we will see in the second next subsection. 
The series of conserved quantities $H_{r,r'}$ are functionally dependent on the previously introduced conserved quantities $Q_{r,r'}$. We give a few examples of how to connect $H_{r,r'}$ with $Q_{r,r'}$ in the thermodynamic limit: 
\begin{equation} \label{eq:Q-to-H}
\begin{split}
 &i^N Q_{0,0}(\theta_0,s_0) = e^{H_{0,0}(i\theta_0,s_0)} \\
 &i^{N-1} Q_{1,0}(\theta_0,s_0) = H_{1,0}(i\theta_0,s_0) e^{H_{0,0}(i\theta_0,s_0)} \\
 &i^{N-2} Q_{2,0}(\theta_0,s_0) = (H_{2,0}(i\theta_0,s_0) + H_{1,0}^2(i\theta_0,s_0)) e^{H_{0,0}(i\theta_0,s_0)} \\
 &i^{N-3} Q_{3,0}(\theta_0,s_0) = (H_{3,0}(i\theta_0,s_0) + 3H_{2,0}(i\theta_0,s_0) H_{1,0}(i\theta_0,s_0) + H_{1,0}^3(i\theta_0,s_0)) e^{H_{0,0}(i\theta_0,s_0)} \\
 &i^N Q_{0,1}(\theta_0,s_0) = H_{0,1}(i\theta_0,s_0) e^{H_{0,0}(i\theta_0,s_0)} \\
 &i^N Q_{0,2}(\theta_0,s_0) = (H_{0,2}(i\theta_0,s_0) + H_{0,1}^2(i\theta_0,s_0)) e^{H_{0,0}(i\theta_0,s_0)} \\
 &i^N Q_{0,3}(\theta_0,s_0) = (H_{0,3}(i\theta_0,s_0) + 3H_{0,2}(i\theta_0,s_0) H_{0,1}(i\theta_0,s_0) + H_{0,1}^3(i\theta_0,s_0)) e^{H_{0,0}(i\theta_0,s_0)}. 
\end{split}
\end{equation}
These relations are obtained by comparing the differential coefficients $\partial^r_{\theta} \partial^{r'}_{s} T(\theta,s)$ at $\theta = \theta_0$, $s = s_0$ expressed in terms of $H_{r,r'}$ with those of $Q_{r,r'}$.

\subsection{Spin-flip invariance}
Now, we restrict our attention to the case $\gamma = \frac{\pi}{l}$. The Hamiltonian \eqref{eq:Hamiltonian} possesses the spin-flip symmetry. Nevertheless, it is known that there exist conserved quantities without spin-flip invariance~\cite{bib:PIP13}. 
We write the Lax operator with the auxiliary space of the spin-$s$ representation as the tensor product of the auxiliary part and the physical part: 
\begin{align}
 &L(i\theta,s) = i\sum_{\alpha \in \{0,z,+,-\}} \sigma^{\alpha} \otimes L^{\alpha}(\theta,s), \\
 &L^0(\theta,s) = \sin\theta \cos(\gamma S^z(s)), \quad
 L^z(\theta,s) = \cos\theta \sin(\gamma S^z(s)), \quad
 L^{\pm}(\theta,s) = \sin\gamma \cdot S^{\mp}(s). \label{eq:Lax_component}
\end{align}
Here we dropped the indices $a,n$. The Pauli matrices $\sigma^{\alpha}$ are twice the $\mathfrak{sl}(2)$ spin operators of spin-$\frac{1}{2}$ representations. It is easy to check that the transfer matrix 
\begin{equation}
 T(i\theta,s) = i^N \sum_{\alpha \in \{0,z,+,-\}} {\rm tr}(L^{\alpha_N}(\theta,s) \cdots L^{\alpha_1}(\theta,s)) 
  \sigma^{\alpha_1} \otimes \cdots \otimes \sigma^{\alpha_N} 
\end{equation}
is spin-flip invariant (SFI) only for (half-)integers $s \leq \frac{l-1}{2}$. The operator part of the transfer matrix consists of the Pauli matrices. Since the spin-flip operator $U_S$ transposes the Pauli matrices, it acts on the transfer matrix as 
\begin{align}
 U_S T(i\theta,s) U_S^{-1} &= i^N \sum_{\alpha \in \{0,z,+,-\}} {\rm tr}(L^{\alpha_N}(\theta,s) \cdots L^{\alpha_1}(\theta,s)) 
  \sigma^{\bar{\alpha}_1} \otimes \cdots \otimes \sigma^{\bar{\alpha}_N} \\
 &= i^N \sum_{\alpha \in \{0,z,+,-\}} {\rm tr}(L^{\bar{\alpha}_N}(\theta,s) \cdots L^{\bar{\alpha}_1}(\theta,s)) 
  \sigma^{\alpha_1} \otimes \cdots \otimes \sigma^{\alpha_N}, \label{eq:flip_T}
\end{align}
where we used the notation $\bar{\alpha}$ defined by 
\begin{equation}
 \bar{0} = 0, \quad \bar{z} = z, \quad \bar{+} = -, \quad \bar{-} = +. 
\end{equation}
From \eqref{eq:Lax_component}, \eqref{eq:flip_T}, we obtain that $U_S T(i\theta,s) U_S^{-1} = T(i\theta,s)$ holds only if the transpose exchanges the $S^{\pm}$-operators. Therefore, the transfer matrix is SFI only for (half-)integers $s \leq \frac{l-1}{2}$. 
The symmetry of the transfer matrix directly determines the symmetry of conserved quantities since they are obtained from the parameter expansion of the transfer matrix. Thus, the conserved quantities $Q_{r,r'}, H_{r,r'}$ associated with complex spin $s$ have no spin-flip invariance.

\subsection{Locality and extensivity of conserved quantities}
Extensivity is a natural property we expect for conserved quantities as thermodynamic variables. 

Extensivity is obtained as a consequence of ``locality'' of an operator~\cite{bib:IMPZ16}. Here we say an operator is local if it is written as a translationally invariant sum of local operators $q^{(r)}$ with support size $r$: 
\begin{equation} \label{eq:local_sum}
 Q = \sum_{x=0}^{N-1} \Pi_x( q^{(r)}\otimes \bm{1}^{\otimes N-r}). 
\end{equation}
$\Pi_x$ is the shift operator $\Pi_x(\sigma^{\alpha_1} \otimes \cdots \otimes \sigma^{\alpha_N}) = \sigma^{\alpha_{1+x}} \otimes \cdots \otimes \sigma^{\alpha_{N+x}}$ under the identification $N+x \equiv x$ due to the periodic boundary condition. 
The question is how much the notion of locality is extended to obtain extensivity. 

The weakest condition to obtain extensivity is called ``pseudolocality''~\cite{bib:IMPZ16}. Pseudolocality is defined through the Hilbert-Schmidt inner product. Within the space $\mathcal{A}_N$ of all translationally invariant traceless operators consisting of $N$ sites, let the traceless deformation of the Hilbert-Schmidt inner product be defined by 
\begin{equation}
 \langle A,B \rangle = \frac{{\rm tr}(A^{\dag} B)}{{\rm tr}(\bm{1}^{\otimes N})} - \frac{{\rm tr}\, A^{\dag}}{{\rm tr}(\bm{1}^{\otimes N})} \frac{{\rm tr}\, B}{{\rm tr}(\bm{1}^{\otimes N})} , \quad
  A,B \in \mathcal{A}_N. 
\end{equation}
Note that $\langle B,A \rangle = \overline{\langle A,B \rangle} \neq \langle A,B \rangle$ for non-Hermitian $A$ and $B$. The deformed norm is defined through the inner product by $||Q|| = \sqrt{\langle Q,Q \rangle}$. The deformed Hilbert-Schmidt inner product satisfy the Cauchy-Schwartz inequality~\cite{bib:IMPZ16}: 
\begin{equation}
 | \langle A,B \rangle | \leq ||A||\; ||B||. 
\end{equation}
We say an operator $Q \in \mathcal{A}_N$ is pseudolocal when the square norm $||Q||^2$ has volume-scaling in the thermodynamic limit $N \to \infty$: 
\begin{equation} \label{eq:extensivity}
 0 < \lim_{N \to \infty} \frac{1}{N} ||Q||^2 = \lim_{N \to \infty} \frac{1}{N} \langle Q,Q \rangle < \infty 
\end{equation}
and finite overlap with at least one local operator $b$: 
\begin{equation}
 \lim_{N \to \infty} \langle b,Q \rangle \neq 0. 
\end{equation}

Slightly stronger locality than pseudolocality is called ``quasilocality''~\cite{bib:IMPZ16}. We call $Q \in \mathcal{A}_N$ a quasilocal operator if there exists, for an operator $Q$ written in the form of translationally invariant sums of local operators and a non-local correction: 
\begin{equation} 
 Q = \sum_{x=0}^{N-1} \sum_{r=2}^N \Pi_x( q^{(r)}\otimes \bm{1}^{\otimes N-r}), 
\end{equation}
a positive $\xi$ such that 
\begin{equation}
 ||q^{(r)}|| \leq C e^{-\xi r} 
\end{equation}
with a constant $C$. Actually, many conserved quantities are known to have quasilocality~\cite{bib:IMP15, bib:P14}. 

Our interest is whether quasilocality and extensivity holds for conserved quantities $Q_{r,r'}(\theta,s)$ by changing the parameters, especially their derivatives $(r,r')$. Among several ways to derive the $N$-dependence of the norms $||Q_{r,r'}(\theta,s)||$~\cite{bib:PPSA14, bib:P14}, we follow the method used in \cite{bib:P14}, which allows us to check the quasilocality as well. 
The case of $(r,r') = (0,1)$ in our notation is discussed in the original paper~\cite{bib:P14}. Since the calculation for an arbitrary set $(r,r')$ is cumbersome, we show the $(r,r') = (0,2)$ and $(1,1)$ cases as representative examples. The arbitrary $(r,r')$ case is also briefly mentioned. 

Due to the parameter expansion of the transfer matrix \eqref{eq:TM_expansion}, we have 
\begin{align}
	Q_{0,2}(\theta_0,s_0) 
	&= \partial_s^2\; T(\theta,s) \Big|_{\theta = \theta_0, s = s_0} \\
	&=  \sum_{\alpha \in \{0,z,+,-\}} {\rm tr}\left\{ \partial_s^2 (L^{\alpha_N}(\theta,s) \cdots L^{\alpha_1}(\theta,s)) \right\} \Big|_{\theta = \theta_0, s = s_0}
  \sigma^{\alpha_1} \otimes \cdots \otimes \sigma^{\alpha_N}. 
\end{align}
By definition, a conserved quantity is written as a translationally invariant summation form:  
\begin{equation}
\begin{split}
	Q_{0,2}(\theta_0,s_0)
 &= 	(\sin\theta)^N (\gamma \cot\theta)^2\sum_{1 \leq x < y \leq N}  \bm{1}^{\otimes x-1} \otimes \sigma^z \otimes \bm{1}^{\otimes y-x-1} \otimes \sigma^z \otimes \bm{1}^{\otimes N-y} \\
	&+ (\sin\theta)^N \sum_{x=0}^{N-1} \sum_{r=2}^N \Pi_x\left( Q_{0,2}^{(r)}(\theta,s) \otimes \bm{1}^{\otimes N-r} \right)
	+ (\sin\theta)^N  \Pi_x \left( P_{0,2}^{(N)}(\theta,s) \right)\\
&+ \text{const}. 
\end{split}
\end{equation}
The first two terms and the constant are obtained from the expectation value on the highest weight vector $|0 \rangle$ in the auxiliary space, while the third term is obtained as the expectation value on the other vectors. For $s_0 = ln$ ($n \in \mathbb{Z}$), the operator $Q_{0,2}^{(r)}(\theta,s_0)$ consists of local operators with support size $r$: 
\begin{equation}
 Q_{0,2}^{(r)}(\theta,s_0) = 
   \begin{cases} 
    \displaystyle
    \gamma \cot\theta \sum_{d=2}^{r-1} q^{(d)}(\theta,s_0) \otimes \bm{1}^{\otimes r-d} \otimes \sigma^z 
    + \gamma \cot\theta \sum_{d=2}^{r-1} \sigma^z \otimes \bm{1}^{\otimes r-d-1} \otimes q^{(d)}(\theta,s_0) 
    + \partial'_s q^{(r)}(\theta,s_0)
    & \displaystyle
    r \leq \frac{N-1}{2} \vspace{3mm}\\
    \displaystyle
    \gamma \cot\theta \sum_{d=2}^{r-1} \sigma^z \otimes \bm{1}^{r-d-1} \otimes q^{(d)}(\theta,s_0) 
    +\partial'_s q^{(r)}(\theta,s_0)
    & \displaystyle
    r \geq \frac{N+1}{2}, 
   \end{cases}
\end{equation}
where $q^{(d)}(\theta,s_0)$ is a local operator with support size $d$: 
\begin{equation}
 q^{(d)}(\theta,s_0) = (\sin\theta)^{-d} 2\gamma \sin\gamma \sum_{\alpha} 
	\langle 1| L^{\alpha_{d-1}}(\theta,s_0) \cdots L^{\alpha_2}(\theta,s_0) |1 \rangle 
	\sigma^- \otimes \sigma^{\alpha_2} \otimes \cdots \otimes \sigma^{\alpha_{d-1}} \otimes \sigma^+. 
\end{equation}
We denote ``the restricted differentiation'' by $\partial'_s$ which acts as 
\begin{equation}
\begin{split}
 \partial'_s q^{(r)}(\theta,s)\Big|_{s=s_0} &= (\sin\theta)^{-r} 2\gamma \sin\gamma \sum_{\alpha} \sum_{k=2}^{{\rm min} (r-1,\frac{N-1}{2})}
	\langle 1| L^{\alpha_{r-1}}(\theta,s) \cdots \partial_s L^{\alpha_k}(\theta,s) \cdots L^{\alpha_2}(\theta,s)|1 \rangle\Big|_{s=s_0} \\
 &\times
	\sigma^- \otimes \sigma^{\alpha_2} \otimes \cdots \otimes \sigma^{\alpha_{r-1}} \otimes \sigma^+ 
\end{split}
\end{equation}
becoming the normal differentiation for $r \leq \frac{N-1}{2}$. Note that these expressions are available only for odd $N$, although the even $N$ case is similarly expressed. 
The non-local correction is also written in terms of the restricted differentiation as 
\begin{align}
 &P_{0,2}^{(N)}(\theta,s) = \partial'_s p^{(N)}(\theta,s), \\
 &p^{(N)}(\theta,s) = (\sin\theta)^{-N} \sum_{\alpha} \sum_{m=1}^{l-1} 
	\langle m| L^{\alpha_N}(\theta,s) \cdots L^{\alpha_2}(\theta,s) \partial_s L^{\alpha_1}(\theta,s)  |m \rangle 
	\sigma^{\alpha_1} \otimes \cdots \otimes \sigma^{\alpha_N}.  
\end{align}
$Q_{0,2}$ is a quasilocal operator if there exists, for the local operators $Q_{0,2}^{(r)}$, a positive $\xi$ such that $||Q_{0,2}^{(r)}|| \leq C e^{-\xi r}$ and the corrections $P_{0,2}^{(N)}$ are exponentially small as the system size grows $||P_{0,2}^{(N)}|| \leq C' e^{-\xi N}$. 
In order to compute the norms $||Q_{0,2}^{(r)}||, ||P_{0,2}^{(N)}||$, we use the real and symmetric matrix $\bm{T}(\bar{\theta},\bar{s};\theta,s)$ defined in \cite{bib:P14}, which originates in ``the double Lax operator''. We also introduce its derivatives $\bm{V}(\bar{\theta},\bar{s};\theta,s), \bm{U}(\bar{\theta},\bar{s};\theta,s)$: 
\begin{align}
	&\bm{T}(\bar{\theta},\bar{s}_0;\theta,s_0) = \sum_{m=1}^{l-1} \left( (\cos(\gamma m))^2 + |\cot\theta|^2 (\sin(\gamma m))^2 \right) |m \rangle \langle m| 
	+ \sum_{m=1}^{l-2} \frac{\sin(\gamma m) \sin(\gamma (m+1))}{2|\sin\theta|^2} \left( |m \rangle \langle m+1| + |m+1 \rangle \langle m| \right), \\
	&\bm{V}(\bar{\theta},\bar{s}_0;\theta,s_0) = \sum_{m=1}^{l-1} \gamma^2 \left( (\sin(\gamma m))^2 + |\cot\theta|^2 (\cos(\gamma m))^2 \right) |m \rangle \langle m|
	+ \sum_{m=1}^{l-2} \frac{2\gamma^2 (\cos(\gamma m))^2 \sin(\gamma(m+1))}{|\sin\theta|^2 \sin(\gamma m)} |m+1 \rangle \langle m|, \\	
	&\bm{U}(\bar{\theta},\bar{s}_0;\theta,s_0) = \sum_{m=1}^{l-1} \gamma^4 \left( (\cos(\gamma m))^2 + |\cot\theta|^2 (\sin(\gamma m))^2 \right) |m \rangle \langle m|
	+ \sum_{m=1}^{l-2} \frac{8\gamma^4 \sin(\gamma m) \sin(\gamma (m+1))}{|\sin\theta|^2} |m+1 \rangle \langle m|.
\end{align}
The square norm of the local operator $q^{(r)}$ is then expressed as 
\begin{equation}
 || q^{(r)}(\theta,s_0) ||^2 
	= (2\gamma \sin\gamma)^2 |\sin\theta|^{-4} \langle 1| \bm{T}(\bar{\theta},\bar{s}_0;\theta,s_0)^{r-2} |1 \rangle
\end{equation}
whose upper bound is evaluated by the leading eigenvalue $\tau_1(\theta,s_0)$ of $\bm{T}(\bar{\theta},\bar{s}_0;\theta,s_0)$. For the spectral parameter $\theta$ in ``the quasilocal strip'' ${\rm Im}\,\theta \in (\frac{l-1}{2} + ln, \frac{l+1}{2} + ln)$ ($n \in \mathbb{Z}$), the eigenvalues are contracting and the largest one $\tau_1$ satisfies $0 < \tau_1 < 1$~\cite{bib:P14}. Note that the conditions ${\rm Im}\,\theta \in (-\frac{1}{2} + ln, \frac{1}{2} + ln)$ also provide another quasilocal strip for $s_0 = l(n + \frac{1}{2})$. Thus we have
\begin{equation}
 || q^{(r)}(\theta,s_0) || \leq C_1 e^{-\xi r}
\end{equation}
with the decay length $\xi = -\frac{1}{2} \log \tau_1 > 0$~\cite{bib:P14}. 
Indeed, the quasilocal operator $Q_{0,1}(\theta,s_0)$ is written by the translationally invariant sums of the exponentially decaying operators $q^{(r)}$ with their support size $r$ ($r=2,\dots,N$)~\cite{bib:P14}. 
The upper bound of the norm $|| \partial'_s q^{(r)} ||$ is also evaluated by $\tau_1$. Using the Cauchy-Schwartz inequality for the Hilbert-Schmidt norm, we obtain
\begin{align}
 || \partial'_s q^{(r)}(\theta,s)\big|_{s = s_0} || \leq 
  \begin{cases}
   \displaystyle C_2 r e^{-\xi r} + \mathcal{O}(e^{-\xi r}) & \displaystyle r \leq \frac{N-1}{2} \\ \\
   \displaystyle C_3 N e^{-\xi r} + \mathcal{O}(e^{-\xi r}) & \displaystyle r \geq \frac{N+1}{2}. 
  \end{cases}
\end{align}
Then we find that the norm of the operators $Q_{0,2}^{(r)}, P_{0,2}^{(N)}$ behave as 
\begin{align}
	&|| Q_{0,2}^{(r)}(\theta,s_0) || \leq
  	\begin{cases}
	   \displaystyle 2C_4 + \mathcal{O}(r e^{-\xi r}) & \displaystyle r \leq \frac{N-1}{2} \\ \\
	   \displaystyle C_4 + \mathcal{O}(N e^{-\xi r}) & \displaystyle r \geq \frac{N+1}{2}, 
	  \end{cases} \\
	&|| P_{0,2}^{(N)}(\theta,s_0) || \leq C_5 N e^{-\xi N} + \mathcal{O}(e^{-\xi N}), 
\end{align}
which indicates that the operator $Q_{0,2}$ is not quasilocal. Moreover, extensivity does not hold for $Q_{0,2}$ since 
\begin{equation}
\begin{split}
	|| Q_{0,2}(\theta,s_0) ||^2 &= N \sum_{r=2}^N || Q_{0,2}^{(r)}(\theta,s_0) ||^2 + 2{\rm Re} || Q_{0,2}^{(r)}(\theta,s_0) ||\;|| P_{0,2}^{(N)}(\theta,s_0) || + || P_{0,2}^{(N)}(\theta,s_0) ||^2 \\
	&\leq N \sum_{r=2}^{N} C_4^2 + 2N^2 \sum_{r=2}^{\frac{N-1}{2}} 2C_4 C_5 N e^{-\xi N} + 2N^2 \sum_{r=\frac{N+1}{2}}^N C_4 C_5 N e^{-\xi N} + C_5^2 N^4 e^{-2\xi N} \\
	&< N^2 C_4^2 + \mathcal{O}(N) 
\end{split}
\end{equation}
not being proportional to the system size $N$. 

For the operator $Q_{1,1}(\theta,s_0)$, we obtain the decomposition: 
\begin{equation}
\begin{split}
  Q_{1,1}(\theta,s_0) &= -\gamma (\sin\theta)^N \sum_{x=1}^N \bm{1}^{\otimes x-1} \otimes \sigma^z \otimes \bm{1}^{\otimes N-x} \\
 &+ \sum_{x=0}^{N-1} \sum_{r=2}^N \Pi_x\left( Q_{1,1}^{(r)}(\theta,s_0) \otimes \bm{1}^{\otimes N-r} \right) + \sum_{x=0}^{N-1} \Pi_x\left( P_{1,1}^{(N)}(\theta,s_0) \right)\\
 &+ \text{const}. 
\end{split}
\end{equation}
Similar analysis reads each component behaving as 
\begin{align}
 &|| Q_{1,1}^{(r)}(\theta,s_0) || \leq C_6 N e^{-\xi r} + \mathcal{O}(r e^{-\xi r}), \\
 &|| P_{1,1}^{(N)}(\theta,s_0) || \leq C_7 N e^{-\xi N} + \mathcal{O}(e^{-\xi N}). 
\end{align}
Thus, the $Q_{1,1}$ is not quasilocal. No extensivity is observed for $Q_{1,1}$ since 
\begin{equation}
\begin{split}
	|| Q_{1,1}(\theta,s_0) ||^2 &= N \sum_{r=2}^N || Q_{1,1}^{(r)}(\theta,s_0) ||^2 + 2{\rm Re} || Q_{1,1}^{(r)}(\theta,s_0) ||\;||  P_{1,1}^{(N)}(\theta,s_0) || + ||  P_{1,1}^{(N)}(\theta,s_0) ||^2 \\
	&\leq N \sum_{r=2}^N C_6^2 N^2 e^{-2\xi r} + 2N^2 \sum_{r=2}^N C_6 C_7 e^{-\xi r} N e^{-\xi N} + C_7^2 N^4 e^{-2\xi N} \\
	&< C_6^2 N^3 \frac{e^{-4\xi}}{1 - e^{-2\xi}} + \mathcal{O}(N^4 e^{-\xi N}), 
\end{split}
\end{equation}
which is not proportional to $N$. 
We found that the conserved quantity $Q_{r,r'}(\theta,s_0)$ for arbitrary $(r,r')$ behaves as 
\begin{equation} \label{eq:N-dep}
	|| Q_{r,r'}(\theta,s_0) ||^2 \sim N^{2r+r'}. 
\end{equation}
Thus, only the operator $Q_{0,1}(\theta,s_0)$ has extensivity, which is also quasilocal. 

The extensivity of another series of conserved quantities $H_{r,r'}$ is showed from direct calculation of their expectation values on ``the Bethe states''. The expectation value of the transfer matrix on the Bethe state has been obtained~\cite{bib:LCN17} as 
\begin{equation} \label{eq:T-function}
	\langle \bm{\lambda}| T(\lambda,s) |\bm{\lambda} \rangle 
	= \mathsf{Q}\left(\lambda + i\gamma(s + \tfrac{1}{2})\right) \mathsf{Q}\left(\lambda - i\gamma(s + \tfrac{1}{2})\right)
	\sum_{m=0}^{l-1} \frac{f(\lambda + i\gamma(s + m))}{\mathsf{Q}(\lambda + i\gamma(s + m + \frac{1}{2})) \mathsf{Q}(\lambda + i\gamma(s + m - \frac{1}{2}))}. 
\end{equation}
A set of Bethe roots $\bm{\lambda} = \{ \lambda_j \}_{j=1,\dots,n}$ characterizes each highest weight eigenstate called the Bethe state. Here we introduced the functions $\mathsf{Q}(\lambda)$ and $f(\lambda)$ defined by 
\begin{align}
	&\mathsf{Q}(\lambda) = \prod_{j=1}^n \sinh(\lambda - \lambda_j), \quad
	f(\lambda) = \left(\sinh\lambda\right)^N. 
\end{align}
For large $N$, the term of $m=0$ becomes dominant for $s = s_0$ around the shift point $\lambda = \frac{i\gamma}{2}$, while the other terms exponentially decay as $N$ grows. Thus, \eqref{eq:T-function} effectively becomes 
\begin{equation} \label{eq:effectiveT}
	\langle \bm{\lambda}| T(\lambda,s) |\bm{\lambda} \rangle 
	\sim f(\lambda + i\gamma s) \frac{\mathsf{Q}(\lambda - i\gamma(s+\frac{1}{2}))}{\mathsf{Q}(\lambda + i\gamma(s-\frac{1}{2}))}. 
\end{equation}
In the thermodynamic limit, the Bethe roots $\lambda_j$ form the Bethe strings by densely distributing along the real axis~\cite{bib:T05}. There are finite $l$ types of Bethe strings for the anisotropy $\gamma = \frac{\pi}{l}$. The string of type $r$ ($r=1, \dots, l-1$) consists of $r$ strings: 
\begin{equation}
	\lambda = \lambda_{\rm R} +  i\gamma \left(m - \frac{r+1}{2}\right),\quad
	m = 1, \dots, r
\end{equation}
with the real center $\lambda_{\rm R} \in \mathbb{R}$, while the Bethe string of type $l$ contains only one string with the shifted center whose imaginary part given by $\frac{i\pi}{2}$. The former types of strings are called ``positive parity'' strings and the latter ``negative parity'' strings.  
By introducing the Bethe string densities $\rho_r(\lambda)$ ($r = 1,\dots,l$), i.e. the density of the Bethe string centers, the logarithmic derivative of \eqref{eq:effectiveT} reads 
\begin{equation} \label{eq:P_TL}
	\lim_{N \to \infty} \frac{1}{N} \langle \bm{\lambda}| H_{r,r'}(\lambda,s) |\bm{\lambda} \rangle 
	= \sum_{r=1}^l \int_{-\infty}^{\infty} d\mu\, h_j^{(r,r')}(\lambda-\mu,s) \rho_j(\mu) 
\end{equation}
for $(r,r') \neq (0,0)$. In the right-hand side, a set of Bethe roots are replaced by a set of Bethe string densities $\{\rho_r(\lambda)\}_{r=1,\dots,l}$ which a set of Bethe roots $\bm{\lambda}$ approaches in the thermodynamic limit. 
The bound-state particle density $h_j^{(r,r')}$ is given by 
\begin{align}
	\widehat{h}_j^{(r,r')}(k,s) = 
	\begin{cases}
	(-ik)^{r-1} \partial_s^{r'} \dfrac{\sinh((l-2s) \frac{\gamma k}{2}) \sinh(j \frac{\gamma k}{2})}{\sinh(\frac{\gamma k}{2}) \sinh(l \frac{\gamma k}{2})} & j = 1,\dots,l-1 \vspace{3mm} \\
	-(-ik)^{r-1} \partial_s^{r'} \dfrac{\sinh(-2s \frac{\gamma k}{2})}{\sinh(l \frac{\gamma k}{2})} & j = l
	\end{cases}
\end{align}
in the Fourier space. Thus, the right hand side of \eqref{eq:P_TL} remains finite in the thermodynamic limit, which indicates the extensivity of $H_{r,r'}$. The list of known locality properties of conserved quantities are provided in Table \ref{tab:locality}.

Note that the SFNI conserved quantities $H_{r,r'}(\lambda,s)$ are linearly independent from those with spin-flip invariance $H_r^{(s)}(\lambda)$ which are associated with (half-)integer spins~\cite{bib:WNBFRC14}. We discuss the functional dependence of the SFNI conserved quantities later in connection with the string-charge duality. 

\begin{table}
\begin{center}
\caption{Locality and extensivity properties of conserved quantities for the gapless $XXZ$ model are listed. Remind that pseudolocaliy is the weakest condition for an operator to be extensive. The spectral parameter $\theta$ is to be in the quasilocal strip. The charge $H_r^{(\frac{1}{2})}(\lambda)$ at the shift point $\lambda = \frac{i\gamma}{2}$ is the local charge which coincides with the Hamiltonian for $r=1$. Although quasilocality of $H_1^{(\frac{l}{2})}(\frac{i\gamma}{2} + t)$ ($t \in \mathbb{R}$) has been proved only at the isotropic point~\cite{bib:IMP15}, we expect the same for the anisotropic case. No classification has been achieved yet for $Q_r^{(\frac{l}{2})}(\theta)$ which are not listed here.} \label{tab:locality}

\vspace{5mm}
\includegraphics[height = 35mm]{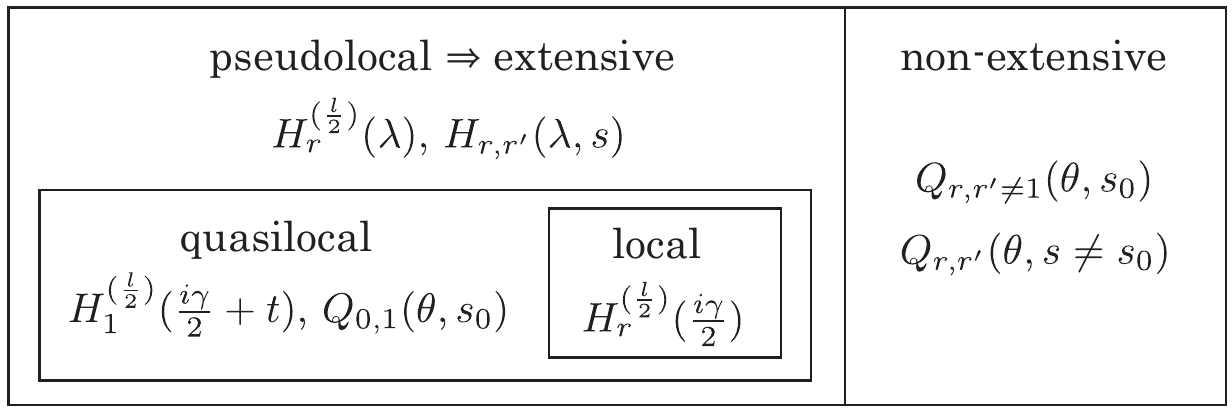}

\end{center}
\end{table}


\section{Generalized Gibbs ensemble}
Unlike non-integrable systems, integrable systems do not thermalize. Instead, they approach to the steady state described by the generalized Gibbs ensemble (GGE)~\cite{bib:RDYO07} that maximize the entropy under the constraint of fixed expectation values of conserved quantities. Later it has been showed that the system locally approaches to the GGE steady state but not as a whole~\cite{bib:BS08, bib:CDEO08}. Finally, the GGE conjecture was formulated in the form 
\begin{equation}
	\lim_{t \to \infty} \frac{\langle \Psi(t)| O_{\rm local} |\Psi(t) \rangle}{\langle \Psi(t)|\Psi(t) \rangle} 
	= {\rm tr} \left( \rho_{\rm GGE} O_{\rm local} \right) 
\end{equation}
was formulated in \cite{bib:CEF12, bib:FE13}. 
The GGE density matrix $\rho_{\rm GGE}$~\eqref{eq:GGE}, whose idea was first introduced in \cite{bib:J57, bib:J57-2}, consists of the macroscopic number of conserved quantities~\cite{bib:RDYO07} where the Lagrangian multipliars $\beta_r$ are determined by the initial condition: 
\begin{equation}
	\frac{\langle \Psi(0)| Q_r |\Psi(0) \rangle}{\langle \Psi(0)|\Psi(0) \rangle} = {\rm tr}\left( \rho_{\rm GGE} Q_r \right), 
\end{equation}
since the expectation value of conserved quantities are invariant under time evolution. (Under the presence of the dynamical symmetry, the discussion must be modified by using the time-dependent GGE introduced in \cite{bib:MBJ19}. )

The question we ask is which conserved quantities form a minimum complete set to describe the GGE. This question has been discussed for the $XXZ$ model by using the string-charge duality~\cite{bib:WNBFRC14, bib:BWFNVC14, bib:IQNB16}. However, we are still far from understanding the general framework to construct the GGE for arbitrary integrable systems. In this section, we show that the GGE for the $XXZ$ model consists of functionally independent conserved quantities. 

\subsection{String-charge duality and linear dependence of conserved quantities}
The string-charge duality provides the relation between the Bethe string densities for the steady state and the expectation values of conserved quantities on the initial state for the $XXZ$ model~\cite{bib:WNBFRC14, bib:BWFNVC14, bib:IQNB16}. As a conserved quantity is invariant under time evolution, its expectation value on the initial state is the same as that on the relaxation steady state. Since the steady state of the $XXZ$ model is characterized by the Bethe string densities~\cite{bib:C16} up to the freedom coming from the $\mathbb{Z}_2$-symmetry, the string-charge duality indicates that a set of conserved quantities which determines all the Bethe string densities is nothing but what constitutes the GGE. 
As the thermodynamic variables, we use the extensive conserved quantities $H_{r,r'}$ rather than non-extensive ones $Q_{r,r'}$ in this section. 

The first trial of finding the relation between the Bethe string densities and the charges has been discussed in \cite{bib:WNBFRC14}, which provides the one-to-one correspondence between all local conserved quantities associated with spin-$\frac{1}{2}$ and the one-string hole density for the zero magnetization initial state. The apparent discrepancy was obtained~\cite{bib:PMWKZT14, bib:WNBFRC14} between the calculation by the GGE with spin-$\frac{1}{2}$ conserved quantities and the results obtained from the microcanonical viewpoint by ``the quench action (QA) method''~\cite{bib:CE13, bib:C16}. The improved GGE, which includes quasilocal conserved quantities associated with (half-)integer spins, has been proposed based on the string-charge duality that connects $l$ string-densities with $l-1$ quasilocal conserved quantities associated with (half-)integer spin~\cite{bib:IQNB16}: 
\begin{equation}
\begin{split}
	\widehat{\rho}_r(k) - \delta_{r,l-1} \widehat{\rho}_l(k) 
	= 2 \cosh\left(\frac{\gamma k}{2}\right) \widehat{H}^{(\frac{r}{2})}_{1}(k) - \widehat{H}^{(\frac{r+1}{2})}_{1}(k) - \widehat{H}^{(\frac{r-1}{2})}_{1}(k). 
\end{split}
\end{equation}
Here we denote the expectation value of conserved quantities in the thermodynamic limit \eqref{eq:P_TL} simply by $H_{r,r'}(\lambda)$. $\widehat{H}_{r,r'}(k)$ is their Fourier transforms. The relation was systematically derived from the $Y$-system~\cite{bib:KSS98}, which is equivalent to the thermodynamic Bethe ansatz (TBA). The mismatch between the number of the string-density functions ($l$) and the conserved quantities ($l-1$) has been explained as a result of the truncation of the $Y$-system~\cite{bib:IQNB16} that occurs for $\gamma = \frac{\pi}{l}$~\cite{bib:KSS98}. The construction of the complete GGE has been achieved in \cite{bib:LCN17} by adding one SFNI conserved quantity $H_{1,1}(\lambda,0)$ as the entry of the string-charge duality: 
\begin{equation} \label{eq:complete_GGE}
	\widehat{\rho}_l(k)
	= -\cosh\left(\frac{\gamma k}{2}\right) \widehat{H}^{(\frac{l-1}{2})}_{1}(k) - \frac{1}{\gamma k} \sinh\left(\frac{\gamma k}{2}\right) \widehat{H}_{1,1}(k,0). 
\end{equation}
However, the question still remaining is why specifically the operator $H_{1,1}(\lambda,s)$ with $s=0$ is chosen to complete the GGE in spite of the existence of infinitely many SFNI conserved quantities. 

We found that the generalized string-charge duality: 
\begin{equation} \label{eq:zastring-charge}
	\widehat{\rho}_l(k)
	= \widehat{G}'^{(r,r')}(k,s) \widehat{H}^{(\frac{l-1}{2})}_{1}(k) + \widehat{G}^{(r,r')}(k,s) \widehat{H}_{r,r'}(k,s)
\end{equation}
for $(r,r') \neq (0,0)$. The main difference from \eqref{eq:complete_GGE} is obtained as the SFNI operator in the last term. This means that any SFNI conserved quantity can determine the string density $\rho_l(\lambda)$. The explicit forms of the functions $G'^{(r,r')}(\lambda,s)$ and $G^{(r,r')}(\lambda,s)$ are, for instance for $H_{1,0}(\lambda,s)$, obtained as 
\begin{equation} 
	\widehat{\rho}_l(k)
	= -\frac{\sinh((l-2s)\frac{\gamma k}{2})}{\sinh((l-2s-1)\frac{\gamma k}{2})} \widehat{H}^{(\frac{l-1}{2})}_{1}(k) + \frac{\sinh(\frac{\gamma k}{2})}{\sinh((l-2s-1)\frac{\gamma k}{2})} \widehat{H}_{1,0}(k,s). 
\end{equation}
For arbitrary choice of $(r,r')$ and $s$, they are 
determined by the relations 
\begin{equation} \label{eq:linear_dep}
\begin{split}
  &\widehat{H}_{1,0}(k,s) = 
 \frac{\sinh((l-2s-1)\frac{\gamma k}{2})}{\sinh((l-2t-1)\frac{\gamma k}{2})} \widehat{H}_{1,0}(k,t)
	-\frac{\sinh((2t-2s)\frac{\gamma k}{2})}{\sinh((l-2t-1)\frac{\gamma k}{2})} \widehat{H}^{(\frac{l-1}{2})}_{1}(k), \\
 &\widehat{H}_{r,2p}(k,s) = 
 (-ik)^{r-1} (-\gamma k)^{2p} \widehat{H}_{1,0}(k,s), \\
 &\widehat{H}_{r,2p-1}(k,s) = 
(-ik)^{r-1} (-\gamma k)^{2p-1} \frac{1}{\sinh((l-2s-1)\frac{\gamma k}{2})} \widehat{H}^{(\frac{l-1}{2})}_{1}(k) \\
 &\hspace{20mm}+ (-ik)^{r-1} (-\gamma k)^{2p-1} \coth((l-2s-1)\tfrac{\gamma k}{2}) \widehat{H}_{1,0}(k,s), 
\end{split}
\end{equation}
which hold for $s$ satisfying ${\rm Im}\,s \in (0,l)$. For $s$ without satisfying this condition, we obtain the similar convolution form. The convolutions are naturally regarded as the generalization of linear combination in the continuous $\lambda$-space. Therefore, the relations indicate that any SFNI conserved quantity is expressed by a generalized linear combination of one SFNI conserved quantity and a spin-flip symmetric term with $H^{(\frac{l-1}{2})}(\lambda)$. 
We can easily check that the SFI conserved quantity $H_r^{(\frac{j}{2})}(\lambda)$ is also linearly dependent on $H_1^{\frac{j}{2}}$ ($j=1,\dots,l-1$) in a very similar way. 

Thus, the known set of conserved quantities $\{ H^{(\frac{j}{2})}_{1}(\lambda) \}_{j=1,\dots,l-1} \cup \{ H_{1,1}(\lambda,0) \}$ that constitutes the complete GGE is equivalent to a set $\{ H^{(\frac{j}{2})}_{1}(\lambda) \}_{j=1,\dots,l-1} \cup \{ H_{r_1,r'_1}(\lambda,s_1) \}$ for any fixed $(r_1,r'_1)$ and $s_1 \in \mathbb{C} \backslash \{ \frac{1}{2}, \dots, \frac{l-1}{2} \}$ in the sense of functional independence (and linear independence here). This is the minimal set to constitute the complete GGE. 
Note that the product of conserved quantities are in general linearly independent but do not cast in the string-charge duality. That is, the complete string-charge duality, and subsequently, the complete GGE consists only of functionally independent conserved quantities. We show that a certain product of conserved quantities is functionally independent from its components rather than linearly independent in the next section. This indicates that the GGE does not contain all linearly independent conserved quantities. 

\subsection{Magnetization as a SFNI conserved quantity}
The characteristic properties of the conserved quantities $H_{r,r'}(\lambda,s)$ associated with complex spin are extensivity and spin-flip non-invariance. On the other hand, the total spin operator, which is not a family of conserved quantities obtained from the transfer matrix associated with (half-)integer spins, is also extensive and pin-flip non-symmetric. This similarity motivates us to expect that the physical meaning of the SFNI conserved quantities is interpreted in terms of the total spin operator. Indeed, the operator $H_{0,1}(\lambda,s_{\rm R} + is_{\rm I})$ coincides with $S^z$ up to multiplicity in the large imaginary-spin limit: 
\begin{equation}
\begin{split}
 \lim_{s_{\rm I} \to \infty} \lim_{N \to \infty} \frac{1}{N} i\pi \langle \bm{\lambda}| H_{1,0}(\lambda,s_{\rm R} + is_{\rm I}) |\bm{\lambda} \rangle
 = \sum_{j=1}^l n_j \int_{-\infty}^{\infty} d\lambda\, \rho_j(\lambda) = \lim_{N \to \infty} \frac{1}{N} \langle \bm{\lambda}| S^z |\bm{\lambda} \rangle. 
\end{split}
\end{equation}
$n_j$ are the lengths of Bethe strings where $n_j = j$ for positive parity strings and $n_j = 1$ for negative parity strings~\cite{bib:T05}. 
Thus, the conserved quantities without spin-flip invariance is regarded as a generating function of the total spin operator.

\subsection{Remarks on Generalized Gibbs ensemble}
The GGE consisting of a set of conserved quantities $\{ H^{(\frac{j}{2})}_{1}(\lambda) \}_{j=1,\dots,l-1} \cup \{ H_{r_1,r'_1}(\lambda,s_1) \}$ ($s_1 \in \mathbb{C} \backslash\{\frac{1}{2}, \dots, \frac{l-1}{2}\}$) never allows a SFNI conserved quantity and its spin reverse to cast at the same time. If this is allowed, we would face the problem of non-diagonalizability of the GGE density matrix, which implies non-existence of the steady state. 
Due to the existence of the SFNI conserved quantity, our GGE correctly describes the non-vanishing spin current, which is remarked in \cite{bib:MPP14, bib:LCN17}. 
Note that, we cannot identify whether we are working on the positive magnetization sector or the negative one only by the information of the Bethe strings since the Bethe ansatz method discusses only either the sector of positive magnetization or that of negative magnetization. 
We must properly choose either of two reference states, i.e., the fully polarized positively magnetized state $|\Omega \rangle$ or the fully polarized negatively magnetized state $|\widetilde{\Omega} \rangle$ according to the initial condition. 

As was showed in \cite{bib:INWCEP15}, the complete GGE of the $XXZ$ model in the gapped regime consists only of the SFI conserved quantities, due to the emergence of an infinite tower of string types which invalidating the notion of string parity. The same Bethe-string structure is obtained at the isotropic point, from which we naturally expect that the complete GGE of the $XXX$ model consists only of the SFI conserved quantities as well. This means that the $XXZ$ model in these regimes never exhibits persistent spin transport, although there exist SFNI conserved quantities which are not necessarily extensive.

\section{Ballistic channels for spin currents}
Another important application of the SFNI conserved quantities is found in the discussion of the non-vanishing spin current. Non-vanishing currents are characteristic phenomena obtained for integrable systems. Ballistic channels of currents are supported by their overlap with conserved quantities~\cite{bib:PI13, bib:SPA09, bib:ZNP97}. 
The existence of ballistic channels for the spin current of the $XXZ$ model has been in discussion for a long time~\cite{bib:BFKS05, bib:Z99} until the discovery of SFNI conserved quantities~\cite{bib:PI13}. 
Besides the spin-flip non-invariance, the key property used in the proof to show the non-vanishing spin current is the system-size dependence of the square norms of conserved quantities which cancels the system-size dependence of the overlap in the thermodynamic limit. Although discussed only for quasilocal conserved quantities, we found that the cancellation occurs also for non-quasilocal conserved quantities. In this section, we discuss which non-quasilocal conserved quantities provide ballistic channels of the spin current and whether they improve the lower bound of the Drude weight. 

\subsection{Lower bound for Drude weight}
Finite Drude weight indicates the existence of non-vanishing DC current. The Drude weight of current is evaluated by the current-current correlation in the framework of the linear reponse theory: 
\begin{equation}
\begin{split}
	D(\beta) &= \lim_{t \to \infty} \lim_{N \to \infty} \frac{\beta}{2Nt} \int_0^t dt'\; \langle J(0), J(t') \rangle_{\beta}.  
\end{split}
\end{equation}
The thermal average $\langle \cdot, \cdot \rangle_{\beta}$ at the temperature $T = \beta^{-1}$ is reduced to the Hilbert-Schmidt inner product in the high temperature limit $T \to \infty$ ($\beta \to 0$). 
The Drude weight is bounded from below by the overlap between the current and conserved quantities~\cite{bib:M69, bib:S71}. The ballistic channel supported by $Q_k$ is obtained as the lower bound for the current-current correlation: 
\begin{equation} \label{eq:ballistic_channel1}
	\lim_{t \to \infty} \lim_{N \to \infty} \frac{1}{2Nt} \int_0^t dt'\; \langle J(0), J(t') \rangle_{\beta}
	\geq	\frac{1}{2N} \frac{| \langle J, Q_k \rangle_{\beta} |^2}{|| Q_k ||^2_{\beta}},  
\end{equation}
which survives in the thermodynamic limit if the ratio $| \langle J, Q_k \rangle_{\beta} |^2 / || Q_k ||^2_{\beta}$ is of the order of the system size. For a finite orthogonal set $\{ Q_k \}$ such that $\langle Q_j, Q_k \rangle = \delta_{j,k} ||Q_k||^2$, a ballistic channel supported by each $Q_k$ does not overlap and thus we have 
\begin{equation} \label{eq:Drude_discrete}
 \lim_{t \to \infty} \lim_{N \to \infty} \frac{1}{2Nt} \int_0^t dt'\; \langle J(0), J(t') \rangle_{\beta}
	\geq	\frac{1}{2N} \sum_k \frac{| \langle J, Q_k \rangle_{\beta} |^2}{|| Q_k ||^2_{\beta}}.  
\end{equation}

The saturation condition for the lower bound is clearly understood by decomposing a current into conserved quantities~\cite{bib:S71}. The idea is to write down a current and any physical quantity, in principle, by a linear combination of conserved quantities $Q_k$ and its non-conserved part $J'$: 
\begin{equation} \label{eq:decomp_currents}
J = \sum_{k} \alpha_k Q_k + J'. 
\end{equation}
The summation is taken over ``a complete set'' of linearly independent conserved quantities~\cite{bib:S71}. Here the completeness means that the dimension of the linear space spanned by the conserved quantities are equal to the dimension of the Hilbert space. That is, we need the same number of conserved quantities as the dimension of the Hilbert space to express an arbitrary physical quantity in the above form. In this context, the product of conserved quantities must be regarded as an independent conserved quantity from its components. This is very much unlike the independence of conserved quantities in the GGE which consists of functionally independent conserved quantities without containing their product. 
The coefficients $\alpha_k$ are, for a finite set $\{ Q_j \}_{j=1,\dots,n}$, to be determined by the inner product $\langle J, Q_k\rangle_{\beta}$~\cite{bib:IMPZ16}: 
\begin{equation}
	\alpha_k = \sum_{j=1}^n \langle J, Q_j \rangle_{\beta} K^{-1}_{jk}. 
\end{equation}
Here we used $\langle J', Q_k \rangle = 0$~\cite{bib:S71}. Thus, we have the current-current correlation in terms of the overlap with the conserved quantities:
\begin{equation}
 \langle J,J \rangle_{\beta} = \sum_{j,k=1}^n \langle J,Q_j \rangle_{\beta} K_{j,k}^{-1} \langle Q_k,J \rangle_{\beta}. 
\end{equation}
If only a subset of conserved quantities is used, the right hand side gives a lower bound. The difference from the previous lower bound \eqref{eq:ballistic_channel1} shows up as the matrix $K = (K_{i,j})_{1 \leq i,j \leq n}$ given by $K_{i,j} = \langle Q_i, Q_j \rangle_{\beta}$, which almost consists of the square norms of conserved quantities but contains their overlap coming from non-orthogonality of conserved quantities. Indeed, the matrix $K$ is invertible only when all $Q_k$ are linearly independent. 

The spin current operator $J_{\rm S}$ is SFAS and, therefore, does not have overlap with the SFI conserved quantities. The finite lower bound of its Drude weight was found to be realized by the quasilocal conserved quantity $Q_{0,1}(\theta,s_0)$ without spin-flip invariance~\cite{bib:P14}: 
\begin{equation}
	D \geq \frac{\beta}{2N}\,
{\rm Re} \int_{\mathcal{D}} d^2\theta\, f(\theta) \langle J_{\rm S},Q_{0,1}(\theta,s_0) \rangle_{\beta}, \quad
	\mathcal{D} = \{\theta\,|\,{\rm Im}\,\theta \in (\tfrac{l-1}{2}, \tfrac{l+1}{2})\}. 
\end{equation}
Since the conserved quantities of the $XXZ$ model have a continuous parameter $\theta$, the summation in \eqref{eq:Drude_discrete} is replaced by the integral. The function $f(\theta)$ solves
\begin{equation}
 \int_{\mathcal{D}} d^2\theta'\, \langle Q_{0,1}(\theta,s_0), Q_{0,1}(\theta',s_0) \rangle_{\beta} f(\theta') = 
\langle Q_{0,1}(\theta,s_0), J_{\rm S} \rangle_{\beta}, 
\end{equation}
which is again almost the inverse of the square norms of the conserved quantities but contains their overlap.

\subsection{Ballistic channels supported by non-quasilocal conserved quantities}
We focus on the high temperature limit $\beta \to 0$ where the thermal average becomes the Hilbert--Schmidt inner product. Now we ask whether non-quasilocal conserved quantities also provide ballistic channels for the spin current. From the relation \eqref{eq:ballistic_channel1}, we expect the existence of ballistic channels supported by non-quasilocal conserved quantities if the ratio $| \langle J_{\rm S}, Q_k \rangle |^2 / || Q_k ||^2$ is of the order of the system size $N$. Since we have already derived the system-size dependence of the denominator in the second section~\eqref{eq:N-dep}, we now compute the system-size dependence of the numerator (current-charge overlaps). Remind that the spin current is defined by 
\begin{equation}
	J_{\rm S} = i \sum_{n=1}^N \left( \sigma_n^+ \sigma_{n+1}^- - \sigma_n^- \sigma_{n+1}^+ \right).   
\end{equation}
Its overlap with the conserved quantity is then calculated as 
\begin{equation}
\begin{split}
	\langle  J_{\rm S}, Q_{r,r'}(\theta,s_0) \rangle
	&= 2^{-N} {\rm tr}(J_{\rm S}^{\dag} Q_{r,r'}(\theta,s_0)) \\
	&= -\frac{i}{4}N \partial_{\theta}^r \partial_s^{r'}  {\rm tr}\left\{ [L^-(\theta,s),\,L^+(\theta,s)] (L^0(\theta,s))^{N-2} \right\} \Big|_{s=s_0}. 
\end{split}
\end{equation}
Taking into account of the vanishing condition for $\partial_{\theta}^r \partial_s^{r'} L^{\alpha}(\theta,s)$ at $s=s_0$, we obtain that only odd $r'$ leads to the finite overlap: 
\begin{equation} \label{eq:overlap1}
\begin{split}
	&\langle J_{\rm S}, Q_{r,2p-1}(\theta,s_0) \rangle \sim N^{r+p} \\
	&\langle J_{\rm S}, Q_{r,2p}(\theta,s_0) \rangle = 0. 
\end{split}
\end{equation}
From \eqref{eq:N-dep} and \eqref{eq:overlap1}, we obtain 
\begin{equation}
	\frac{| \langle J_{\rm S}, Q_{r,2p-1}(\theta,s_0) \rangle |^2}{|| Q_{r,2p-1}(\theta,s_0) ||^2} \sim N
\end{equation}
for any positive integer $p$. Therefore, any non-quasilocal conserve quantity $Q_{r,r'}(\theta,s_0)$ with odd $r'$ provides a ballistic channel for the spin current. 

The next question is whether the non-quasilocal conserved quantities $Q_{r,2p-1}$ improve the lower bound of the Drude weight. Actually, it seems that a continuous set of quasilocal conserved quantities already saturates the lower bound~\cite{bib:P14}: 
\begin{equation}
	D \geq \frac{\beta}{2N}\,
	{\rm Re} \int_{\mathcal{D}} d^2\theta\, f(\theta) \langle J_{\rm S}, Q_{0,1}(\theta,s_0) \rangle_{\beta}
	= 1 - \frac{l}{2\pi} \sin\left(\frac{2\pi}{l}\right)	
\end{equation}
in comparison with the result obtained by the thermodynamic Bethe ansatz~\cite{bib:Z99}. This contradictory-looking fact is explained if the non-quasilocal conserved quantities $Q_{r,2p-1}$ are written as the convolution, i.e. the continuum generalization of the linear combination of the quasilocal ones $Q_{0,1}$ in the thermodynamic limit. 
As the SFAS operator, the spin current is decomposed as 
\begin{equation} \label{eq:decomp_current}
J_{\rm S} = \frac{1}{2}\int_{\mathcal{D}} d^2\theta\, \sum_{r,p} \alpha_{r,2p-1}(\theta) (Q_{r,2p-1}(\theta,s_0) - U_S Q_{r,2p-1}(\theta,s_0) U_S^{-1}) + J_{\rm S}'. 
\end{equation}
The convolution relation 
\begin{equation} \label{eq:linear_dep_Q}
\begin{split}
	&Q_{r,2p-1}(\theta,s_0) = \int_{\mathcal{D}} d^2\theta'\,  K_{r,2p-1}(\theta-\theta',s_0) Q_{0,1}(\theta',s_0) 
\end{split}
\end{equation}
reads the spin current in terms of the quasilocal conserved quantities $Q_{0,1}$: 
\begin{equation}
\begin{split}
 &J_{\rm S} 
 = \frac{1}{2}\int_{\mathcal{D}} d^2\theta\, \widetilde{\alpha}_{0,1}(\theta) (Q_{r,r'}(\theta,s_0) - U_S Q_{r,r'}(\theta,s_0) U_S^{-1}) + J'_{\rm S}, \\
 &\widetilde{\alpha}_{1,0}(\theta) = \sum_{r,r'} \int_{\mathcal{D}} d^2\theta'\, \alpha_{r,r'}(\theta') K_{r,r'}(\theta'-\theta,s_0),  
\end{split}
\end{equation}
which indeed indicates that the spin current is purely supported by the quasilocal conserved quantities. 
We are able to convert the convolution relations \eqref{eq:linear_dep_Q} into the functional relations among $H_{r,r'}$: 
\begin{equation} \label{eq:linear_dep_H}
\begin{split}
 &e^{H_{0,0}(i\theta,s_0)} H_{r_1,r'_1}^{n_1}(i\theta,s_0) H_{r_2,r'_2}^{n_2}(i\theta,s_0) \dots \\
 &= \int_{\mathcal{D}} d^2\theta'\, K^{\{n_1,n_2,\dots\}}_{\{(r_1,r'_1),(r_2,r'_2),\dots\}}(\theta-\theta',s_0) e^{H_{0,0}(i\theta',s_0)} H_{0,1}(i\theta',s_0). 
\end{split}
\end{equation} 
Due to the existence of the factor $e^{H_{0,0}}$, the relation \eqref{eq:linear_dep_H} indicates that the product of $H_{r,r'}(\lambda,s_0)$ is not written by the linear combination of $H_{0,1}(\lambda,s_0)$ unless $H_{0,0}(\lambda,s_0)$ is proportional to identity. This is consistent with our statement in the previous section that the GGE for the gapless $XXZ$ model consists of functionally independent conserved quantities, i.e. the GGE does not contain all linearly independent conserved quantities.

\section{Concluding remarks}
In this paper, we discussed nonequilibrium behaviors of the gapless $XXZ$ model brought by SFNI conserved quantities. First, we showed that the GGE is given by a set of functionally independent conserved quantities. We derived the generalized string-charge duality which connects all SFNI conserved quantities by convolution, which is the continuum generalization of linear combination. The physical meaning of the SFNI conserved quantities is also provided. We found that the total spin $S^z$ operator is obtained from the large imaginary spin limit of $H_{1,0}$. 
The second result is the existence of ballistic channels for the spin current supported by non-quasilocal conserved quantities. In the derivation, we used the system-size dependence of the square norms of conserved quantities and the overlap between the current and the conserved quantities. The saturation of the lower bound for the Drude weight implies that non-quasilocal conserved quantities are expressed by the convolution of quasilocal ones. We obtain that this convolution relation is consistent with the statement that the GGE consists of functionally independent conserved quantities. 

As we mentioned in the abstract, a general framework to construct the GGE has not yet been found. 
Although we found that the GGE for the gapless $XXZ$ model consists of a set of functionally independent conserved quantities, it has been proposed that the product of conserved quantities must be added to correctly describe the steady state of the Lieb--Liniger model with finite repulsive coupling, if one starts from the initial state with long-range correlation~\cite{bib:GA15}. We leave it as a future work to answer the question which kind of independence is required for conserved quantities to completely describe the steady state of arbitrary integrable systems. As is known as an example in which the GGE consisting only of local conserved quantities fails~\cite{bib:Y16}, the attractive Lieb--Liniger model would be a good starting point. 

The second question we did not answer in this paper is what is the completeness of conserved quantities which characterize the steady state. Originally the string-charge duality has been derived based on the $Y$-system~\cite{bib:IQNB16, bib:KSS98}, which is equivalent to the thermodynamic Bethe ansatz. Our next project would be to put the SFNI conserved quantities in the framework of the $Y$-system. There exists the discussion to construct the Baxter's $Q$-operator associated with the complex-spin auxiliary space~\cite{bib:BLMS10} and we expect this would help to proceed this project. 

The third problem we are interested in is how the SFNI conserved quantities lose their contribution to the long-time steady state at the isotropic point (and also in the gapped regime). If naively considered, the SFNI conserved quantities are no more extensive in these regimes and they have no contribution in the thermodynamic limit. Contrarily, the Drude weight has a finite value in the isotropic case at zero temperature~\cite{bib:BFKS05}, which implies the existence of the SFNI conserved quantities valid in the thermodynamic limit. 

In the classical case, it is known that ``quasi-integrable systems'' possess chaotic structure although having the singularity confinement property~\cite{bib:KMT16}. By analyzing these systems, we expect to see how nonequilibrium behavior changes as the system becomes toward integrable. We leave this question as a future work.


\section*{Acknowledgements}
The author acknowledges F. G\"{o}hmann, A. Kl\"{u}mper, B. Pozsgay, and N. Tsuji for helpful discussions. C. M. is supported by JSPS Grant-in-Aid, No. 11J10068, Japan and JST CREST, No. JPMJCR14D2, Japan.

\bibliographystyle{abbrv}
\bibliography{references}

\end{document}